# Driving carbon nanotube to rotate by diamond wedges at room temperature

Jiao Shi [1], Aiqin Wang [1], Kun Cai [2,*]

[1] College of Water Resources and Architectural Engineering, the Northwest A&F University, Yangling 712100, China
[2] Centre for Innovative Structures and Materials, School of Engineering, the RMIT University, Melbourne 3001, Australia
*Corresponding author's email address: kun.cai@rmit.edu.au (K. C.)

**Abstract**: A rotary nanomotor made of carbon nanostructures is introduced here. Through a rotationally symmetrical layout of diamond wedges (or needles) outside of a carbon nanotube and with the [100] direction of diamond along the tube's axial direction, the wedge needle tips can drive the nanotube to rotate at gigahertz frequency at room temperature. During thermal vibration, some of the atoms in the nanotube collide with the needle tips. The tips provide the atoms with continuous repulsion during collision. The tangential component of the repulsion force produces a moment onto the nanotube about the tube axis. Consequently, the nanotube is driven to rotate by the moment. The rotor reaches stable rotation when the concentric outer tubes provide equivalent resistance against the repulsion from needle tips. Molecular dynamics simulation results indicate that the stability of the needle tips influences the rotational frequency of the rotor. Potential fabrication of such a nanomotor is illustrated with consideration of miniaturization. The rotary motor can act as an engine in a nanomachine.

## 1. Introduction

Carbon has many natural allotropes with different configurations at nanoscale, e.g., fullerene (zero-dimension) [1, 2], nanotube (one-dimension) [3], graphene (two-dimension) [4, 5], and diamond (three-dimension). Except for diamond being formed by $sp^3$-$sp^3$ bonds between neighboring carbon atoms, for the low dimensional carbon materials, they are formed by $sp^2$-$sp^2$ bonds. According to the peculiar electron structures of carbon atoms, many artificial carbon nano materials have been developed [6-9].

During the past four decades, physical properties of carbon materials have been widely investigated. Owing to the very high strength of the carbon-carbon bonds, carbon materials show excellent mechanical properties. For example, diamond is one of the hardest natural materials. The in-plane modulus of graphene or nanotube reaches ~1000 GPa with a strength over 100 GPa [10, 11]. Hence, nano-devices made from the carbon nano materials have high stability under mechanical loading. Another interesting property of graphene layers of multi-walled carbon nanotubes (CNTs) is that the inter-layer/shell friction is extremely low [12-14] due to the fact that each atom has an anti-bonding delocalized electron. Low friction means small energy dissipation during the relative sliding [15, 16] between components in such nanodevices as oscillators [15-18], motors [19-24], and bearings [25-27].

As a representative motion device, a nanomotor can convert different types of energy into mechanical motion [19-22, 26, 28]. Nanomotors have been considered a key component of a nanomachine or a sensor using bottom-up approaches for decades. In particular, researchers have been trying to fabricate nanomotor from CNTs in recent years. For example, Fennimore et al. [19] fabricated a nanoscale electromechanical actuator, in which multi-walled CNTs acted as the shaft. On the shaft, a metal plate was attached and driven to rotate under an external electric field. The size of the nanomotor was ~300 nm which was far greater than that of a molecular motor. Wang et al. [21] presented a model of a nanomotor, in which conducting blades were attached to a carbon nanotube shaft. In an external electric field, the blades were driven to rotate by periodic charging and discharging of the blades. Barreiro *et al.* [26] built an artificial nanomotor from CNTs. The short outer tube attached to cargo could be driven to move on the long inner tube, which had a thermal gradient along the tube axis. Recently, Cai *et al.* [22] discovered that the inner tube in double-walled CNTs had a stable unidirectional rotation in the fixed outer tube in a thermostat with constant temperature. Furthermore, they [23] designed a thermally driven rotary nanomotor with controllable rotation. The method for measuring the rotation of the thermally driven rotary nanomotor was also proposed [29].

In this study, we propose a new method to drive a CNT to rotate. In the new nanomotor from carbon nanostructures, CNTs act as a rotor or stator and diamond needles as an actuator (Figure 1). Kinetic energy for the rotation of the rotor is from the environment, e.g., a thermostat. The validity of the nanomotor model is verified using numerical simulations, and the influences of the stability and geometry of diamond needles on the rotation of the rotor are evaluated.

## 2. Model and Methodology

2.1 Model of a rotary nanomotor

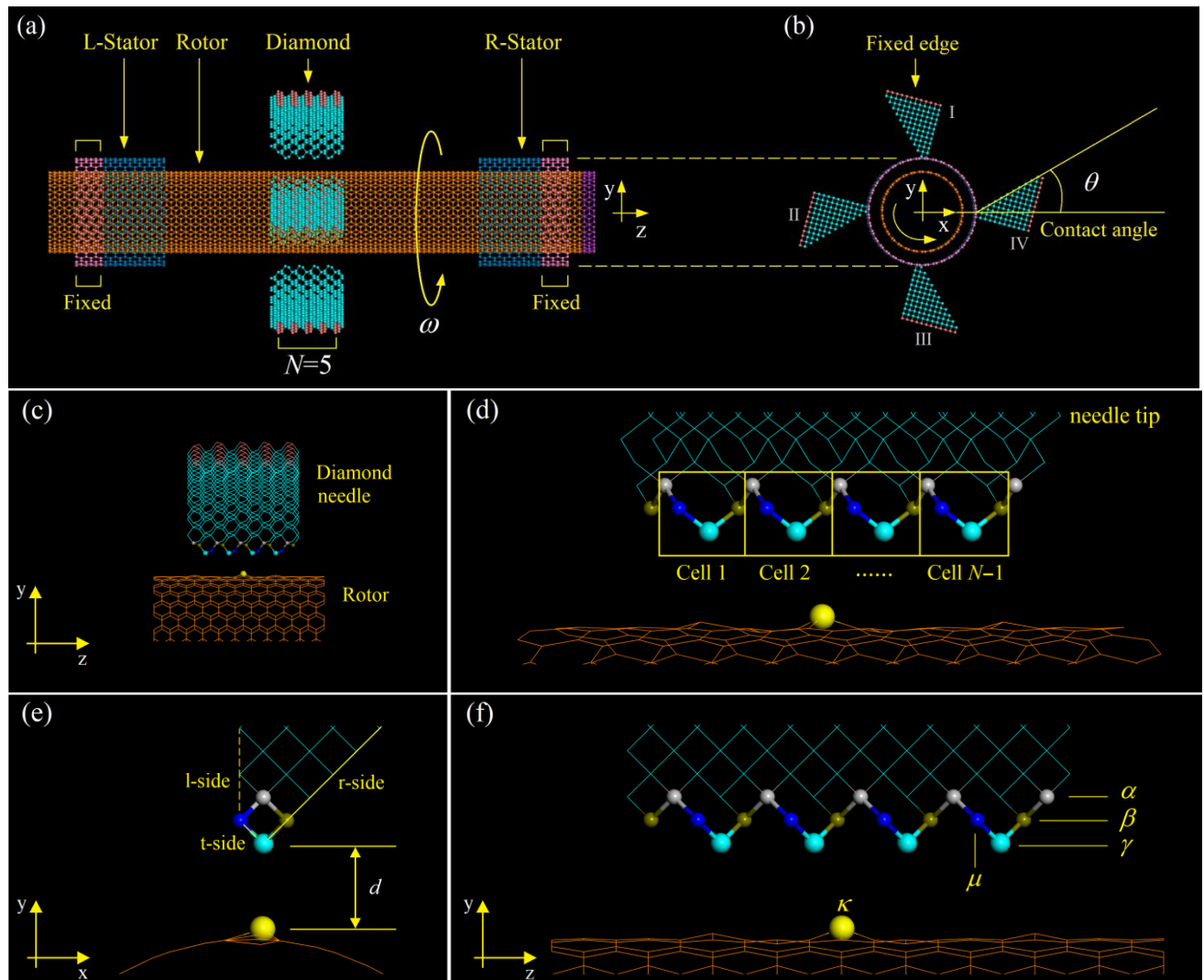

Figure 1 Geometry of the rotary nanomotor from CNTs and diamond. **(a)** Side and **(b)** axial views of motor. The nanomotor consists of three components: two stators from the same CNT, one rotor from a slim CNT and four wedge diamond needles with the normal of their [100] lattice plane along the axial (z-) direction of the rotor. Each diamond needle tip has *N* layers along z-direction. The diamond needles are rotationally

symmetric outside the rotor and have the same contact angle $\theta$. **(c)** Local configuration of a diamond needle. **(d)** Oblique-view of needle tip with $N-1$ periodic cells. **(e)** Each needle tip has three sides, i.e., left (l-side), right (r-side), and tip sides (t-side), which only contain blue atoms and white atoms. "$d$" is the radial distance between the tip and the rotor. **(f)** Four types of atoms at the needle tip, labeled "$\alpha$", "$\beta$", "$\gamma$", and "$\mu$", which have a strong interaction with an atom (e.g., yellow atom $\kappa$) on the CNT rotor at finite temperature. "$\omega$" is the rotational frequency of the rotor when driven by the diamond needles.

Figure 1 gives the geometric configuration of the nanomotor made from CNTs and diamond needles. For the diamond needle with $N$ layers, it has only $N-1$ tip atoms, i.e., $\gamma$ atoms (Figure 1f). At finite temperature, the atoms on the CNT rotor have a large amplitude of thermal vibration, which leads to $d<0.3$nm. Hence, the needle tips have stronger repulsion onto the rotor at atom $\kappa$ (Figure 1f). The four rows of atoms at each needle tip are significant to the repulsion. The four rows of atoms are labeled as "$\alpha$", "$\beta$", "$\gamma$", and "$\mu$". Within each cell at a needle tip (Figure 1d), each $\alpha$ atom has four $sp^3$ bonds, and each $\beta$ atom has reshaped three $sp^2$ bonds. Both $\gamma-$ and $\mu-$ atoms are unsaturated, e.g., each $\gamma$-atom is bonded with a $\mu$ atom and a $\alpha$ atom, and each $\mu$ atom is bonded with $\gamma$ and $\beta$ atoms. Both the l-side and r-side have unsaturated carbon atoms.

## 2.2 Mechanism for rotation of rotor driven by diamond needles

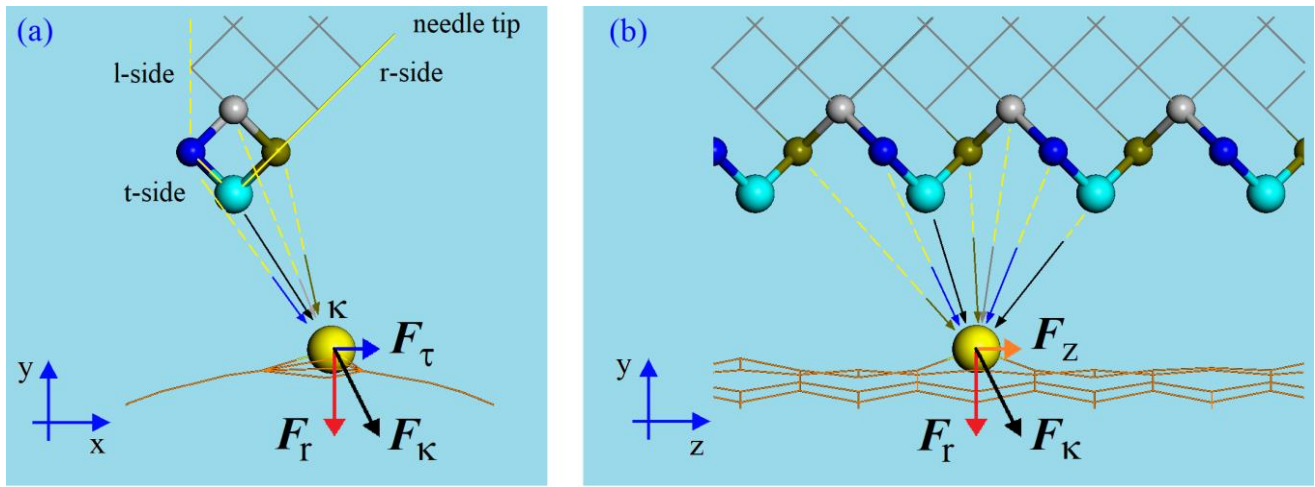

Figure 2  Schematic of free body diagram of an atom ($\kappa$) on a rotor subjected to repulsion from a diamond needle tip. (a) Within the cross-section across atom $\kappa$, the components of $F_\kappa$ are $F_\tau$ and $F_r$. (b) In the longitudinal cross-section, $F_\kappa$ has two components, $F_z$ and $F_r$.

At finite temperature, the atoms on the rotor have thermal vibration. The relative positions of the rotor and needle tips change during vibration. When the value of $d$ is less than 0.34 nm but higher than 0.2 nm, the atoms at the needle tips provide nonbonding repulsion onto the rotor. For example, Figure 2, atom $\kappa$ near a diamond needle tip is subjected to the strong repulsion from the tip atoms and much weak attraction from other atoms on the needle. The resultant force on atom $\kappa$ is $F_\kappa$, which has three components, i.e., $F_\tau$ along the tangent direction of the rotor, $F_z$ along the tub axis of rotor, and $F_r$ towards the tube axis. Hence, the four vectors from the $i$th needle satisfy the following equation, i.e.,

$$F_\kappa^{(i)} = F_\tau^{(i)} + F_r^{(i)} + F_z^{(i)} \quad (i=1, 2, 3, 4). \tag{1}$$

When the rotor is confined within two stators, $F_r$ leads to a breathing vibration of the rotor, $F_\tau$ provides a non-zero moment onto the rotor, and $F_z$ provides translational acceleration of the rotor. Among the three components, $F_\tau$ is kept relatively stable compared to the other two components, and the rotational frequency of the rotor caused by the $F_\tau$ is,

$$\omega(t) = \int_0^t \frac{1}{J_z} \left[ \sum_{i=1}^4 M\left(F_\tau^{(i)}\right) - \sum_{j=1}^2 M\left(F_f^{(j)}\right) \right] ds, \tag{2}$$

where $M(F)$ represents the moment produced by force vector $F$ about tube axis. $F_f$ is the friction force from both stators, and $J_z$ is the transient rotary inertia of the rotor about its tube axis (z-axis) at time t. $F_f$ increases with the rotational frequency of the rotor. Only when $F_\tau$ and $F_f$ provides the same but opposite values of moment on the rotor, does the rotor have stable rotational frequency.

2.3 Methodology of analysis

To provide a quantitative analysis of the rotational frequency of the rotor in the nanomotor shown in Figure 1, the molecular dynamics method is adopted in the open source code LAMMPS [30]. The interaction among neighboring atoms are estimated by the AIREBO potential [31], which can reflect bonding interaction via a REBO item, nonbonding interaction by a Lennard-Jones part, and the dihedral effect by a torsion part, i.e.,

$$\begin{cases} P = P_{REBO} + P_{Torsion} + P_{L-J} \\ P_{REBO} = \sum_i \sum_{j(j>i)} \left[ V_{ij}^R(r_{ij}) - b_{ij} V_{ij}^A(r_{ij}) \right] \\ P_{Torsion} = \frac{1}{2} \sum_i \sum_{j(j\neq i)} \sum_{k(k\neq i,j)} \sum_{l(l\neq i,j,k)} w_{ij}(r_{ij}) \cdot w_{jk}(r_{ij}) \cdot w_{kl}(r_{ij}) \cdot V_{Torsion}(\omega_{ijkl}) \\ P_{L-J} = \sum_i \sum_{j(j>i)} 4\varepsilon \left[ \left(\frac{\sigma}{r_{ij}}\right)^{12} - \left(\frac{\sigma}{r_{ij}}\right)^6 \right] \end{cases} \tag{3}$$

where $V_{ij}^R$ and $V_{ij}^A$ in the short-range REBO potential illustrate impulsion and attraction with respective to atoms $i$ and $j$, respectively, and $b_{ij}$ illustrates the many-body term. $r_{ij}$ is the distance between atoms $i$ and $j$. $P_{Torsion}$ describes the potential due to the change of the dihedral angle $\omega$ of the atoms $i$, $j$, $k$, and $l$. The bond weights, $w_{ij}$, are in the interval [0, 1], $P_{L-J}$ is the Lennard-Jones potential [32] with $\sigma=0.34$ nm, $\varepsilon=2.84$ meV, and 1.02 nm as cutoff.

For a system in non-equilibrium state, the potential energy of the system varies continuously. The variation of the potential energy (VPE) of the system depends on the configuration of the system. Considering Eq.(3), the VPE can be expressed as

$$\text{VPE} = P_t - P_{t0}, \tag{4}$$

where $P_t$ and $P_{t0}$ are the potential energy of the system at time $t$ and $t_0$, respectively. In this study, $t_0$ is set to zero.

3. **Numerical simulations and discussion**

To verify our prediction, i.e., that the nanotube can be driven to rotate by the diamond needles, numerical simulations are performed. In simulations, the two stators in the nanomotor shown in Figure 1 are from the same CNT (20, 20) with a length of ~2.21 nm and the rotor from CNT (15, 15) with a length of ~13.77 nm. The axial distance between both stators is 8.00 nm. The radial distance between each needle tip and the rotor is $d=$~0.3 nm (Figure 1e). Owing to the saturated carbon atoms in diamond having a slight thermal vibration at finite temperature, the fixed edge of the needle tip has no chance to influence the rotor when their distance is larger than 1.0nm. To avoid the boundary effect, the length of the r-side of a needle tip is set to be larger than 1.2 nm in the present study.

After building the model according to the specified geometry, the system is reshaped by minimization of the potential energy with the steepest-decent method. The time-step for integration is set to be 0.001 ps. The initial velocities of atoms in the system satisfy Gaussian distribution, and the temperature of the system is 300 K. Furthermore, the outer edges of both stators are fixed with their degrees of freedom, as are the edges of the four diamond needles. A Nose-Hoover thermostat [33, 34] is adopted to control the temperature. The dynamic response of the system is investigated within the NVT ensemble (N is the total number of atoms, V is the volume, and T is the temperature of system. They keep unchanged with T=300 K.

3.1 Effect of *N* on rotor rotation

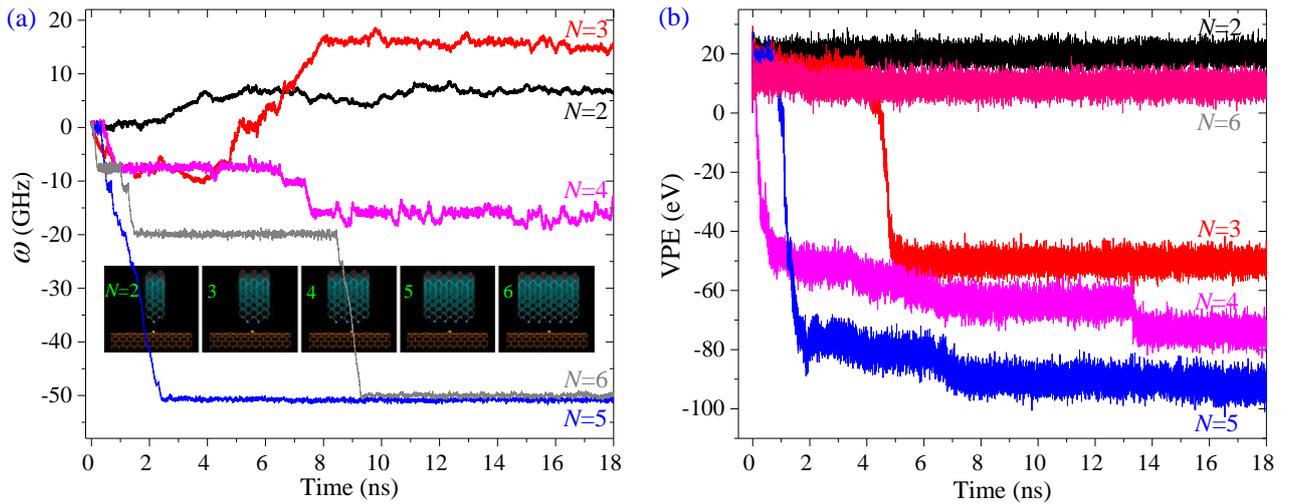

Figure 3  History curves of $\omega$ and VPE of system with $\theta=0°$. (a) Rotational frequency of rotor driven by diamond needles with different numbers of layers. (b) VPE of system.

First, we investigate the effect of the needle tip's width on the stable rotational frequency of the rotor. Figure 3(a) illustrates the histories of $\omega$, the rotational frequency of the rotor, during the first 18ns when the rotor is driven by the needles with different values of $N$ and $\theta=0°$ for all cases. After no more than 10ns does the rotor have stable rotation. For example, when the rotor is driven by the diamond needles with $N=2$, $\omega$ finally reaches ~7 GHz. A positive value of $\omega$ means that the rotor rotates in an anti-clockwise direction about the z-axis (Movie 1). Observing the black curve in Figure 3(b), the VPE of the system with respect to $N=2$ does not decrease further after 12ns. Accordingly, we conclude that the system is in an equilibrium state, and that the rotor has stable rotation.

Similarly, one can find that the final rotational direction is also anti-clockwise when the rotor is actuated by the diamond needles with $N=3$. However, the final stable value of $\omega$ with respect to $N=3$ is more than twice that with respect to $N=2$. However, at initial stage, e.g., [0, 500] ps, the rotor with respect to $N=3$ rotates in a clockwise direction rather than in an anti-clockwise direction. However, the rotational speed of the rotor reduces to be zero before 5ns and further increases to be positive, which means the rotational direction changes to anti-clockwise. The related (red) curve shown in Figure 3(b) also shows the variation process, i.e., VPE drops suddenly from 4 ns to 5 ns. After 5ns, the VPE decreases slightly and finally becomes stable.

When the rotor is driven by the needles with $N≥4$, the rotor rotates always in a clockwise direction (Movies 2). If $N=4$, the rotational frequency of the rotor changes frequently during the first 8 ns. According to the VPE curve in pink shown in Figure 3(b), i.e., the VPE does not drop further, $\omega$ fluctuates near 17GHz after 14ns. When $N=5$, the rotor turns to stable rotation soon after ~2.5ns of acceleration (blue curve in Figure 3a). However, the system enters a stable state after 8ns according to the VPE curve shown in Figure 3(b). One can also find that the final values of $\omega$ of the rotor with respect to $N=5$ and 6 are slightly different demonstrating that the stable rotational frequency of the rotor does not increase further if $N$ becomes larger.

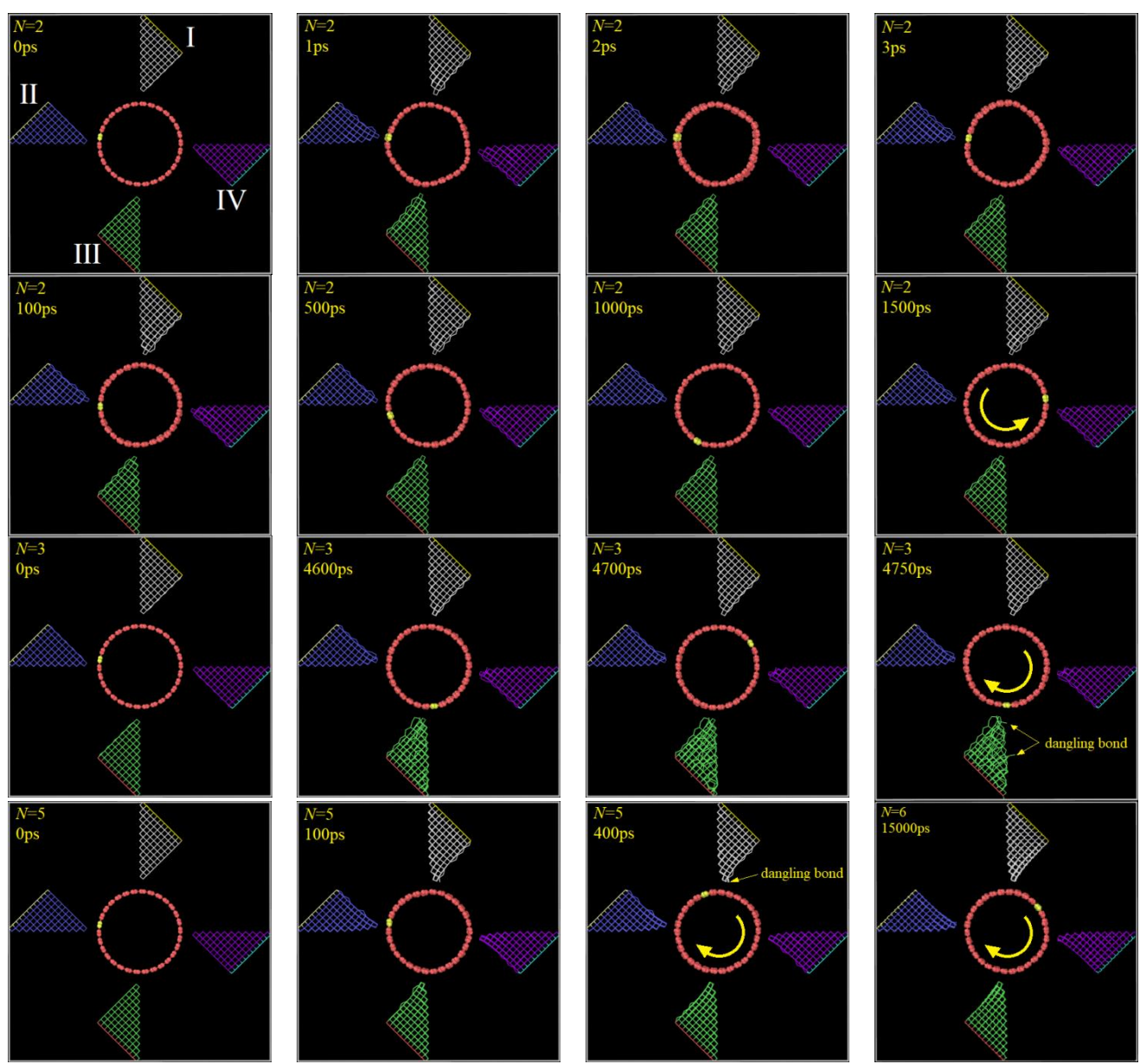

Figure 4 Representative snapshots of (the cross-section of) the nanomotor with different values of $N$.

Previously, we predicted that the unidirectional rotation of the rotor is caused by collision between the rotor and the diamond needle tips. From the first row of Figure 4 with respect to $N=2$, the rotor exhibits drastic breathing deformation. In addition, the needle tips also have obvious deformation as compared to the initial configuration (0 ps). The deformation of both the rotor and the needle tips has two causes. One is their interaction, i.e., repulsion, due to $d$ being less than $\sigma$. The other reason is that the components of the nanomotor are not in equilibrium states without enough relaxation in the thermostat. After 100 ps of relaxation, the rotor starts to rotate as can be seen in the second row of Figure 4. In fact, each needle tip with $N=2$ has only one cell (Figure 1d or the inset of Figure 3a), which provides very weak repulsion on the rotor. Meanwhile, the t-side of the needle tip has only two unsaturated atoms, and is only constrained by an $\alpha$ atom and a $\beta$ atom with three bonds (Figure 1f). Hence, the t-side can move relatively easier than the t-side of the needle tip with $N>2$. At the t-side, the bond lengths are harder to change than the bond angles. Hence, the t-side with $N=2$ can have a larger rotational angle. Because of the rotation of the t-side, the needle tip provides positive moment on the rotor. Therefore, the rotor rotates in an anti-clockwise direction.

When $N>2$, each needle tip has more than one cell, and the t-side has more $\gamma/\mu$ atoms. The bond between $\gamma$ and $\mu$ atoms only has smaller rotational angular compared to that with $N=2$. Moreover, the $C_\gamma$-$C_\mu$ bonds are laid out along the axial (z-) direction (Figure 1c), and form into a wall along the z-direction. The wall together with the rest of the carbon atoms in the needle tip provides a clockwise moment on the rotor during their collision. Hence, the rotor will rotate in the clockwise direction.

The fact that the rotational direction of the rotor with respect to $N=3$ changes from clockwise to anti-clockwise is because of the breakage and rearrangement of a bond at tip III (Figure 1b) near 4800 ps. Owing to the rearrangement, the tip provides an anticlockwise moment onto the rotor. Hence, the rotational direction of the rotor with respect to $N=3$ is not exceptional.

One can also find the breakage and rearrangement of C-C bonds at one or more needle tips in other cases. For example, when $N=5$, tip I has broken bonds, and one of them even is in dangling state (Movie 2). However, the rotational direction of the rotor does not change after the breakage because the tip has more $C_\gamma$-$C_\mu$ bonds.

3.2 Effect of $\theta$ on rotor rotation

From the previous free body diagram shown in Figure 2, the contact angle between a needle tip and the rotor also influence the resultant repulsion on the rotor. To demonstrate the influence, we choose models with both $N=3$ and $N=5$. In each model, $\theta$, the contact angle, is set to be 0°, 10°, 20°, 22.5°, 30° and 45°. The historical curves of $\omega$ with respect to different values of $\theta$ are given in Figure 5. Obviously, the curves of $\omega$ demonstrate that the stable rotational frequency of the rotor depends on the contact angle.

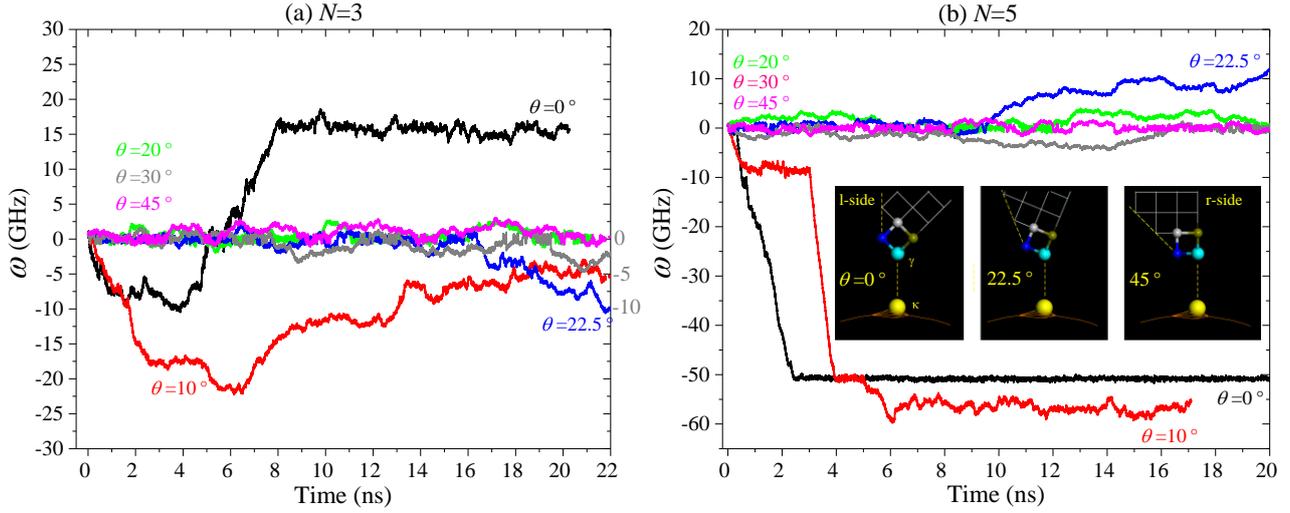

Figure 5  History curves of the rotational frequency of rotor driven by diamond needles with different numbers of layers and different contact angles. (a) $N=3$, and (b) $N=5$.

When observing the dynamic process of the system, one can find that at least one needle tip incurs severe damage during collision with the nanotube rotor (Figure 6). As one of the tips is damaged, its t-side exhibits a weaker reaction against the impulsion from the nanotube. Under the repulsion from the rest of the needles, the rotor has to adjust its equilibrium position, e.g., get closer to the damaged needle tip. In this condition, the reaction between the rotor and the needle tips becomes weaker. Hence, the rotor is difficult to actuate to have a stable rotation within a period of time, e.g., 10 to 20 ns. This is why we find that the rotor, which is driven by needles with either $N=3$ or $N=5$, and $\theta=20°$, $30°$ or $45°$, has no stable rotation within 20 ns (Figure 5). For example, for the system with $N=5$ and $\theta=45°$, tip I has been severely damaged at 18 ns. There are nano-pores (see the shadows in Figure 7; also see Movie 3) and the rest of the atoms near the t-side form into honeycomb cells. The quasi-two-dimensional network formed by the cells and the nano-pores cannot provide a strong reaction to the rotor.

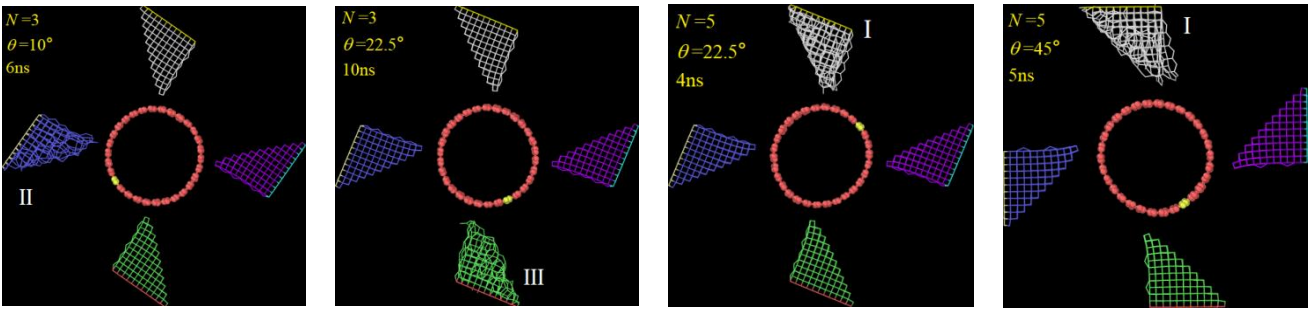

Figure 6  Four snapshots of the system with different damaged diamond needles.

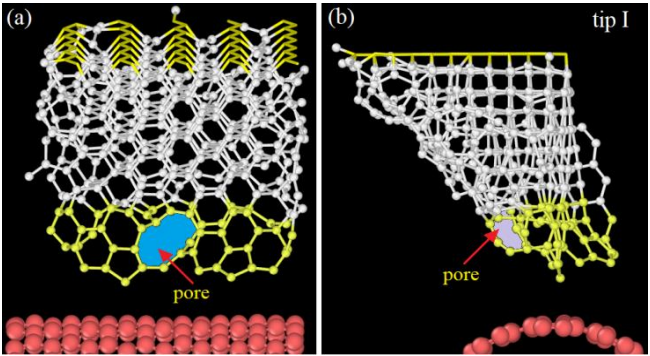

Figure 7  Local configuration of diamond needle tip I with $N=5$ and $\theta=45°$ at 18ns.

Except for the above cases in which the rotor is failed to be driven to rotate, in the rest of the cases the rotor has obvious rotational frequency after 20 ns. For instance, when the rotor is driven by the needles with $N=3$ and $\theta=10°$, it rotates in a clockwise direction (the red curve in Figure 5a). The rotational speed increases in the first 6ns and then decreases to ~5 GHz after 20 ns. The reason is that tip II exhibits a damage smaller than that of tip I with $N=5$ and $\theta=45°$.

Owing to the damage of a needle tip, some other phenomena can be observed. For example, if the rotor is driven by the needles with $N=3$ and $\theta=22.5°$, it rotates in a clockwise direction after 16 ns, and the rotational frequency is over 10 GHz at 22 ns (the blue curve in Figure 5a and the second snapshot in Figure 6). When $N=5$ and $\theta=22.5°$, the rotor's rotational direction is in anti-clockwise after 9 ns, and $\omega$ reaches ~10 GHz soon after that (the blue curve in Figure 5b and the third snapshot in Figure 6). When $N=5$ and $\theta=10°$, the historical curve of the rotor's rotational frequency has three steps, e.g., $\omega=$~8 GHz before 3 ns, ~-50 GHz between 4 ns and 5 ns, and finally ~56 GHz after 12 ns. According to the results, we conclude that the stability of the needle tips depends on the contact angle, and influences the rotor rotation.

## 4. Conclusions

Through a rotationally symmetric layout of four diamond needles outside a carbon nanotube, the nanotube can be driven to rotate at gigahertz frequency at room temperature. The mechanism is that the atoms in the nanotube have drastic thermal vibration, which leads to collision between the nanotube and diamond needle tips. A tangential component of their interaction produces an axial moment on the nanotube, resulting in rotation of the nanotube. The feasibility of the carbon nanomotor is verified by molecular dynamics simulations. According to the obtained results, the following conclusions are drawn:

(a) Without damage of the diamond needle tips, because of large variations of the bond angles at the needle tip with $N=2$, the rotational direction of the rotor is opposite that of the rotor driven by the needles with $N>2$.

(b) The stable rotational frequency of the rotor does not increase further when $N>5$ at 300 K.

(c) The rotational direction of the rotor may change because of the breakage and rearrangement of bonds at diamond needle tips.

(d) When a needle tip incurs severe damage during collision with the nanotube rotor, the reaction between the rotor and the needle tips becomes weaker, and it is difficult for the rotor to achieve stable rotation within few nanoseconds.

(e) The stability of the needle tips depends on the contact angle. Larger contact angle leads to easier breakage of the needle tips, and may stop the rotation of the rotor.

By using the rotary nanomotor as an engine, we can design a nanomachine/nanocar working at finite temperature. The diamond needle tips can be assembled on the frame of the machine and the rotor can directly act as a wheel or drive an external rotary component to move. As the dimensions of the nanomotor are less than 15 nm, miniaturization of a nanomachine can be realized.

**Supporting materials**

Movies

    Movie 1—Motor with N=2 & sita=0 degree at 300K during [4, 4.3]ns.avi
    Movie 2—Motor with N=5 & sita=0 degree at 300K during [0, 0.5]ns.avi
    Movie 3—Motor with N=5 & sita=45 degree at 300K during [0.05, 0.35]ns.avi

**Acknowledgements**

The authors declare no competing financial interests. Financial support from the National Key Research and Development Plan, China (Grant No.: 2017YFC0405102) is acknowledged.